\RequirePackage{fix-cm}
\documentclass[twocolumn,epjc3]{svjour3}  
\smartqed  

\usepackage{amsfonts}
\usepackage{amssymb}
\usepackage{amsmath}
\usepackage{cite}
\usepackage{cancel}
\RequirePackage{graphicx}
\usepackage{dcolumn}
\usepackage{bm}
\usepackage{slashed}
\usepackage{lineno}
\usepackage{IEEEtrantools}
\usepackage{csquotes}
\usepackage{makecell}
\usepackage{multirow}
\usepackage{cancel}
\usepackage[colorlinks = true,
            linkcolor = blue,
            urlcolor  = blue,
            citecolor = blue,
            anchorcolor = blue]{hyperref}
\usepackage{tikz,xcolor}
            
\DeclareOldFontCommand{\tt}{\normalfont\ttfamily}{\mathtt}
\definecolor{lime}{HTML}{A6CE39}
\DeclareRobustCommand{\orcidicon}{
	\begin{tikzpicture}
	\draw[lime, fill=lime] (0,0) 
	circle [radius=0.16] 
	node[white] {{\fontfamily{qag}\selectfont \tiny ID}};
	\draw[white, fill=white] (-0.0625,0.095) 
	circle [radius=0.007];
	\end{tikzpicture}
	\hspace{-2mm}
}

\foreach \x in {A, ..., Z}{\expandafter\xdef\csname orcid\x\endcsname{\noexpand\href{https://orcid.org/\csname orcidauthor\x\endcsname}
			{\noexpand\orcidicon}}
}

\journalname{Noname}
\begin{document}

\title{Uncovering tau leptons-enriched semi-visible jets at the LHC
}


\author{Hugues Beauchesne\thanksref{e1,addr1}\orcidA{}
        \and
        Cesare Cazzaniga\thanksref{e2,addr2}\orcidB{}  
        \and 
        Annapaola de Cosa\thanksref{e3,addr2}\orcidC{} 
        \and
        Caterina Doglioni\thanksref{e4,addr3}\orcidE{} 
        \and 
        Tobias Fitschen\thanksref{e5,addr3}\orcidD{} 
        \and 
        Giovanni Grilli di Cortona\thanksref{e6,addr8,addr9}\orcidF{} 
        \and
        Ziyuan Zhou\thanksref{e7,addr2,addr7}\orcidG{} 
}

\thankstext{e1}{e-mail: beauchesneh@phys.ncts.ntu.edu.tw}

\thankstext{e2}{e-mail: cesare.cazzaniga@cern.ch}

\thankstext{e3}{e-mail: adecosa@phys.ethz.ch}

\thankstext{e4}{e-mail: caterina.doglioni@cern.ch}

\thankstext{e5}{e-mail: tobias.fitschen@cern.ch}

\thankstext{e6}{e-mail: grillidc@lnf.infn.it}

\thankstext{e7}{e-mail:
zizhou@whu.edu.cn}

\institute{Physics Division, National Center for Theoretical Sciences, Taipei 10617, Taiwan \label{addr1}
\and 
ETH Z\"urich, Institute for Particle Physics and Astrophysics, CH-8093 Z\"urich, Switzerland \label{addr2}
\and 
Univeristy of Manchester, School of Physics and Astronomy, M13 9PL Manchester, United Kingdom \label{addr3}
\and  
Istituto Nazionale di Fisica Nucleare, Sezione di Roma, Piazzale A. Moro 2, I-00185 Roma, Italy \label{addr8}
\and  
Istituto Nazionale di Fisica Nucleare, Laboratori Nazionali di Frascati, C.P. 13, 00044 Frascati, Italy \label{addr9}
\and 
School of Physics and Technology, Wuhan University, Wuhan 430072, China \label{addr7}
}

\date{Received: date / Accepted: date}

\maketitle

\begin{abstract}
This Letter proposes a new signature for confining dark sectors at the Large Hadron Collider. 
Under the assumption of a QCD-like hidden sector, hadronic jets containing stable dark bound states could manifest in proton-proton collisions. 
We present a simplified model with a $Z'$ boson yielding the production of jets made up of dark bound states  and subsequently leading to the decays of those that are unstable to $\tau$ leptons and Standard Model quarks. 
The resulting signature is characterised by non-isolated $\tau$ lepton pairs inside semi-visible jets. We estimate the constraints on our model from existing CMS and ATLAS analyses. We propose a set of variables that leverage the leptonic content of the jet and exploit them in a supervised jet tagger to enhance the signal-to-background separation. 
Furthermore, we discuss the performance and limitations of current triggers for accessing sub-TeV $Z'$ masses, as well as possible strategies that can be adopted by experiments to access such low mass regions.
We estimate that with the currently available triggers, a high mass search can claim a $5 \sigma$ discovery (exclusion) of the $Z'$ boson with a mass up to 4.5~TeV (5.5~TeV) with the full Run~2 data of the LHC when the fraction of unstable dark hadrons decaying to $\tau$ lepton pairs is around $50\%$, and with a coupling of the $Z'$ to right-handed up-type quarks of 0.25. Furthermore, we show that, with new trigger strategies for Run~3, it may be possible to access $Z'$ masses down to 700 GeV, for which the event topology is still composed of two resolved semi-visible jets.
\end{abstract}


\section{Introduction}
\label{intro}

Despite the great success of the Standard Model, there is much evidence that it is incomplete. The presence of dark matter in the universe (DM)~\cite{Bertone:2004pz,Planck2018} is one of the most tantalising hints for physics beyond the horizon of the Standard Model (SM). The lack of experimental confirmation of the most simple hypotheses sought so far~\cite{PhysRevLett.39.165,Jungman:1995df} connects to the idea of new physics belonging to sectors of matter barely accessible to Standard Model particles, or \emph{dark sectors}, that has not been fully probed at collider experiments. Several Hidden Valley models~\cite{Strassler:2006im} predict the existence of hidden dark sectors that are only weakly coupled to the SM. 
These models can rather easily accommodate DM candidates, identifying them with stable particles within the dark sector (see e.g. Ref. \cite{SpierMoreiraAlves:2010err,Lee:2015gsa,Okawa:2016wrr,Hochberg:2018vdo,Beauchesne:2018myj,Bernreuther:2019pfb,Beauchesne:2019ato}).  
Assuming that the dark sector is strongly coupled, and that the confinement scale of the theory is sufficiently small, the mixture of stable and unstable dark bound states coming from the hadronization of dark quarks generates semi-visible jets at collider experiments~\cite{Cohen:2015toa,Cohen:2017pzm,Beauchesne:2017yhh,Beauchesne:2018myj,Bernreuther:2019pfb,Beauchesne:2019ato,Cazzaniga:2022hxl}. 
These appear as visible jets of particles in the detectors, with missing transverse momentum ($\cancel{E}_{\text{T}}$) aligned with the direction of one of the jets. 
Fully hadronic semi-visible jets signatures have been proposed in ~\cite{Cohen:2015toa,Cohen:2017pzm,Beauchesne:2017yhh,Bernreuther:2019pfb} and searched for by the CMS and ATLAS experiments in s- and t-channel production modes~\cite{CMS:2021dzg,ATLAS-CONF-2022-038}. 
Signatures with semi-visible jets enriched in leptons from dark photon decays have been recently proposed in~\cite{Cazzaniga:2022hxl}. 
In this Letter we present a simplified model developed following~\cite{Fox:2011qd}, in which a $Z'$ mediator acts as a messenger between the visible and dark sectors, coupling to SM quarks and $\tau$ leptons, enabling the decay of dark bound states to $\tau$ lepton pairs and leading to a $\tau$-enriched semi-visible jets (SVJ$\tau$) signature. 
The effective fraction of invisible particles inside the jet is expected to be larger with respect to the fully hadronic case due to the neutrinos from $\tau$ decays, thus making it more difficult to reconstruct the mediator mass. 
Electrons and muons are also likely to be produced within the jet from leptonic $\tau$ decays,  leading to atypical signatures with high neutrino and lepton multiplicities within a hadronic jet. We find that present searches have limited sensitivity to this novel signature, and propose a SVJ$\tau$ tagger that leverages the features from $\tau$ leptons inside the jet. 

\section{Model setup}
\label{sec:Model}
The general interactions between the $Z'$ boson and SM fermions is given by:
\begin{eqnarray}\label{eq:LagrangianUFO}
\mathcal{L}_{\text{SM}}  \supset  &- Z'_\mu \bar{u}_i \gamma^\mu(g^{u_R}_{ij} P_R + g^{u_L}_{ij} P_L) u_j \nonumber \\
                                 & - Z'_\mu \bar{d}_i \gamma^\mu(g^{d_R}_{ij} P_R + g^{d_L}_{ij} P_L) d_j \nonumber \\
                                  &- Z'_\mu \bar{e}_i \gamma^\mu(g^{e_R}_{ij} P_R + g^{e_L}_{ij} P_L) e_j.
\end{eqnarray}
An explanation of how such couplings can be embedded in gauge invariant effective theories can be found in~\cite{Fox:2011qd}, while~\cite{Altmannshofer:2014cfa} gives an example of UV completion in terms of vector fermions. 
We introduce dark quarks $q_{vi}$ charged under the non-abelian gauge group $SU(N)'$. Such hidden sector particles interact with the $Z'$ boson via the general interactions:
\begin{eqnarray}
  \mathcal{L}_{q_v} \supset -Z'_\mu \bar{q}_{vi} \gamma^\mu (g^{q_{vR}}_{ij} P_R + g^{q_{vL}}_{ij} P_L) q_{vj}.
\end{eqnarray}
The dark sector will contain pseudo-scalar mesons, which can be expressed via the matrix $\Pi = \pi_0 \mathbb{I} + \pi_a t_a$, where the $t_a$ are the broken flavour symmetry generators. 
Assuming the coupling matrices to SM fermions to be diagonal, their decay widths to SM fermions $f_i$ will be:
\begin{equation}\label{eq:DecayWidth}
\begin{aligned}
  \Gamma_{\pi_a \to f_i \bar{f}_i} =\;\frac{N_c F_\pi^2}{32\pi}(\Delta^{f}_{ii}\,\Delta^{q_v}_a)^2 \frac{m_{\pi^a}m_{f_i}^2}{M_{Z'}^4}\sqrt{1 - \frac{4m_{f_i}^2}{m_{\pi_a}^2}},
\end{aligned}
\end{equation}
where $N_c$ is the number of colours of $f_i$, $F_\pi$ is the dark pion decay constant, $M_{Z'}$ is the $Z'$ boson mass, and we define $\Delta^{f}_{ii} = g_{ii}^{f_R} - g_{ii}^{f_L}$, $\Delta^{q_v}_a = g^{q_{vR}}_a - g^{q_{vL}}_a$. The dark couplings are defined via $g^{q_{vR}} = g^{q_{vR}}_0 \mathbb{I} + g^{q_{vR}}_a t_a$ and $g^{q_{vL}} = g^{q_{vL}}_0 \mathbb{I} + g^{q_{vL}}_a t_a$. \\
More specifically, we will present results for the following benchmark. 
We assume two flavours of dark quarks that transform under the fundamental representation of an $SU(3)'$ group. 
The couplings are taken as:
\begin{equation}\label{eq:Benchmark}
  \begin{aligned}
  &g^{d_R}_{ij} = g^{d_L}_{ij} = g^{u_L}_{ij} = g^{e_L}_{ij} = 0, \quad g^{u_R}_{ij} = g_u \delta_{ij}, \\ &g^{e_R}_{ij} = g_{\tau}\delta_{i3}\delta_{j3}, \quad g^{q_{vR}}_{ij} = g^{q_{vL}}_{ij} = g_{q_v} \delta_{ij}.
  \end{aligned}
\end{equation}
\begin{figure}
\includegraphics[scale=0.37]{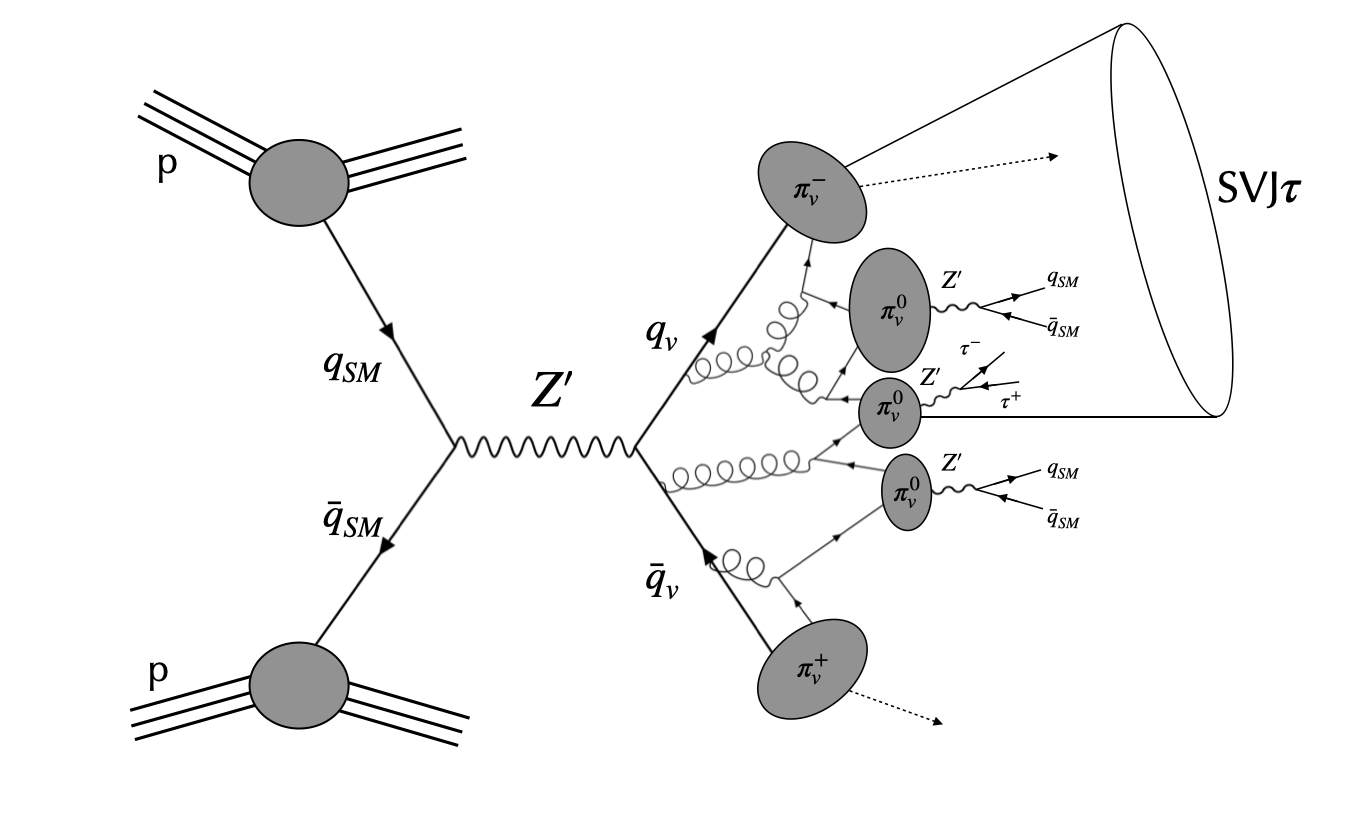}
\caption{\label{fig:s-channel SVJtau diagram}S-channel production of semi-visible jets with $\tau$ leptons from dark hadrons decays.}
\end{figure}
While studying the collider signatures, we will assume the presence of some small deviations of the $g^{q_{vL}}_{ij}$ and $g^{q_{vR}}_{ij}$ couplings to break the $SU(2)$ flavour symmetry between the two dark quarks and change the fraction of stable mesons produced. 
The choice of small $Z'$ couplings to electrons and muons for our benchmark is driven by the strong constraints from high mass di-lepton searches \cite{CMS:2018ipm,ATLAS:2019erb}. 
A coupling to light leptons for a $\sim 1 $ TeV $Z'$ boson mass would be constrained to be much smaller than the $\tau$ lepton coupling. 
Finally, the coupling $g_{q_v}$ has been fixed to 0.6, such that the narrow width approximation holds when one takes into account the number of flavours and colours in the dark sector as in~\cite{CMS:2021dzg}. 
We choose to vary $g_u/g_{\tau}$ implying scanning over different values for the branching ratio of unstable dark sector bound states to $\tau$ leptons ($\text{BR}_{\tau}$).  \\ \\
As a major phenomenological consequence, the proposed simplified model allows for s-channel production of semi-visible jets from $pp \to Z' \to q_v \bar{q}_v$ with an enriched $\tau$ lepton content, originating from dark bound states, as illustrated in Figure~\ref{fig:s-channel SVJtau diagram}. 
Therefore, its final states will be characterised by a large multiplicity of $\tau$ leptons decaying hadronically or leptonically, and whose decay products will be clustered in jets with aligned missing momentum from the stable dark bound states (SVJ$\tau$ signature). 
In this study we only consider dark bound states decaying promptly, since the expected $c \tau_{\text{min}}$ can be so small that displacements between the production and decay points of unstable dark hadrons cannot be resolved. 

\section{Monte Carlo Simulations}
\label{sec:MC simulations}
Samples containing signal events, including initial- and final-state radiation and underlying event, have been produced with {\tt Pythia8.307}~\cite{Bierlich:2022pfr}. 
The Hidden Valley module~\cite{Carloni:2011kk,Carloni:2010tw} in {\tt Pythia8} was used to simulate the showering, hadronization, and decays in the dark sector. 
Detector effects are simulated using {\tt Delphes3}~\cite{deFavereau:2013fsa}.
Following previous studies~\cite{CMS:2021dzg}, anti-$k_T$ jets~\cite{Cacciari:2008gp,Cacciari:2011ma} with \mbox{$R = 0.8$} (AK8 jets) are clustered requiring a minimum $p_{\text{T}}$ of 200 GeV. 
Even if no double hadronization process is expected in case of dark bound states decaying to $\tau$ leptons, the choice of AK8 jets in our study is mainly motivated by a better $Z'$ boson mass resolution.
Since the dark bound states can have a large $\Delta R$ separation from the jet axis due to showering and hadronization in the dark sector, $\tau$ lepton decay products can be produced at relatively large angles, and a large jet radius can help containing them. 
All the samples have been normalised to the LO cross-section prediction computed with the {\tt MadGraph5\_aMC@NLO}~\cite{Alwall:2014hca} event generator using the {\tt NN23LO1} parton distribution functions ~\cite{Ball:2013hta} from the {\tt Lhapdf}~\cite{Buckley:2014ana} repository. \\ \\
For the SVJ${\tau}$ s-channel signal process, $50 \cdot 10^3$ events have been generated scanning a mass range for $M_{Z'}$ between 0.5 TeV and 6 TeV. 
The number of stable and unstable dark hadrons produced in the dark hadronization process can vary according to the details of the dark sector. 
In previous literature\cite{Cohen:2015toa,Cohen:2017pzm}, an effective invisible fraction parameter has been defined as $r_{\text{inv}}=  \langle {N_{\text{stable}}}/{(N_{\text{stable}} + N_{\text{unstable}})} \rangle$, where  $N_{\text{stable}}$ is the number of stable dark hadrons, and $N_{\text{unstable}}$ is the number of those that are unstable, decaying back to SM. This invisible fraction allows to  capture variations in the details of the hidden sector, and define the parameter space of a semi-visible jet signature in terms of amount of missing transverse momentum $\cancel{E}_{\text{T}}$ \footnote{Other operative definitions of this parameter are being studied, see e.g. \cite{Albouy:2022cin}}.  The quantity $\cancel{E}_{\text{T}}$ is computed from the transverse component of the sum of the momenta of all detected particles in an event~\cite{deFavereau:2013fsa}, and can be used to quantify the amount of transverse momentum missing due to invisible particles.
The higher $r_{\text{inv}}$, the more the final state topology will resemble a mono-jet signature, where $\cancel{E}_{\text{T}}$ recoils against initial-state radiation jets. 
Due to the effect of the neutrinos produced from $\tau$ leptons originating from DM bound states, the effective invisible fraction inside the jets is expected to increase. 
We estimate this increase due to neutrinos by adding a term to the above definition of invisible fraction  $ r^{\nu}_{\text{inv}} = r_{\text{inv}} + \langle f_{\nu}\rangle \times \text{BR}_{\tau} \times (1-r_{\text{inv}})$, where  $\langle f_{\nu}\rangle $ represents the invisible component carried by neutrinos averaged over all possible $\tau$ lepton decays, and estimated to be~$\sim 0.45$. 
Thus, for a fixed $\text{BR}_{\tau}$, it is possible to determine the expected experimental signature for SVJ$\tau$ in terms of the alignment of the missing transverse momentum vector $\cancel{E}_{\text{T}}$ with the jets.
In this work, we scan over $\text{BR}_{\tau}$ and fix $r_{\text{inv}} = 0.3$, such that $r^{\nu}_{\text{inv}}$ is smaller then 0.7. This regime can be probed by a SVJ-like search with event selections similar to those performed by ATLAS and CMS~\cite{CMS:2021dzg,ATLAS-CONF-2022-038}. \\ 
The dark pseudo-scalar ($\pi_v$) and vector ($\rho_v$) meson masses can differ according to the non-perturbative dynamics of the hidden sector, even for mass-degenerate dark quarks. 
Following a similar approach as in \cite{Cazzaniga:2022hxl}, we employed the lattice QCD fits in~\cite{Albouy:2022cin} to predict the masses of dark vector mesons $m_{\rho_v}$ from the input ratio $m_{\pi_v}/\Lambda_v$, where $\Lambda_v$ ({\tt HiddenValley:Lambda}) is the dark confinement scale , which fixes the overall mass scale for the dark bound states.  
As a benchmark, $\Lambda_v$ has been set to 10 GeV and the ratio $m_{\pi_v}/\Lambda_v = 0.8$, therefore fixing $m_{\pi}$ ({\tt 4900111:m0, 4900211:m0}) at 8 GeV and $m_{\rho_v}$ ({\tt 4900113:m0, 4900213:m0}) at $\sim 25$~GeV. 
With these dark QCD parameter settings, the $\rho_v \to \pi_v \pi_v$ decay is open, and we assume $100 \%$ branching ratio for this internal decay in the dark sector. 
As a consequence, only unstable pseudo-scalar dark mesons are allowed to decay back to the SM, and within the chosen parameter space, their decays will be mainly into $\tau$ leptons and $c$ quarks. \\
In summary, there are four parameters sensitive to the details of the dark sector: the confinement scale $\Lambda_v$, the pseudo-scalar mass ratio $m_{\pi_v}/\Lambda_v$, the invisible fraction $r_{\text{inv}}$, and the branching ratio to $\tau$ leptons $\text{BR}_{\tau}$. 
Moreover, there are two portal parameters: the $Z'$ pole mass $M_{Z'}$, and the coupling to dark quarks $g_{q_v}$. 
In order to probe the impact of enhanced $\tau$ lepton production, we scan over the branching $\text{BR}_{\tau} (0.15, 0.3, 0.55, 0.7)$ which is equivalent to varying the ratio $g_{u}/g_{\tau}$. Thus, fixing $g_{u}/g_{\tau}$, $\text{BR}_{\tau}$ is fixed, and the signal cross-section is determined for a given benchmark $g_u~=~0.25$ .\\
Following the same approach as in \cite{Cazzaniga:2022hxl}, we have considered the same background processes as in the CMS semi-visible jets search~\cite{CMS:2021dzg}. 
All the background samples have been generated at LO with the {\tt MadGraph5\_aMC@NLO} event generator using the {\tt NN23LO1} parton distribution functions from the {\tt Lhapdf} repository. The evolution of the parton-level events and hadronization are performed with {\tt Pythia8}. 
The QCD sample ($4.5 \cdot 10^7$ events) has been produced requiring a generator level cut on the leading parton jet transverse momentum $p_{\text{T}} > 200$ GeV. 
The QCD background is particularly relevant due to the large cross-section and the possibility of mis-reconstruction of the jet momentum leading to additional missing momentum aligned with the jet axis. Moreover, jets initiated from bottom quarks mimic the characteristic leptonic content of signal jets, due to b-flavored hadrons decays.
The $\text{t} \bar{\text{t}} + \text{jets}$ inclusive sample ($5 \cdot 10^7$ events) has been generated with up to two additional partons. 
This background mainly becomes relevant when the top quarks are highly boosted, and therefore the W boson decay and the b-initiated jet are merged into a larger jet. The electroweak inclusive backgrounds $\text{Z}(\nu \bar{\nu}) + \text{jets} $ and $\text{W} (\ell \nu) + \text{jets}$ ($2.5 \cdot 10^7$ events each) have been produced with a generator level cut $H_{\text{T}} > 100$ GeV and including up to three additional partons in the matrix element. 

\section{Trigger and search strategies}
\label{sec:3}
\begin{figure}[!thbp]
 \centering
 \includegraphics[width=.5\textwidth]{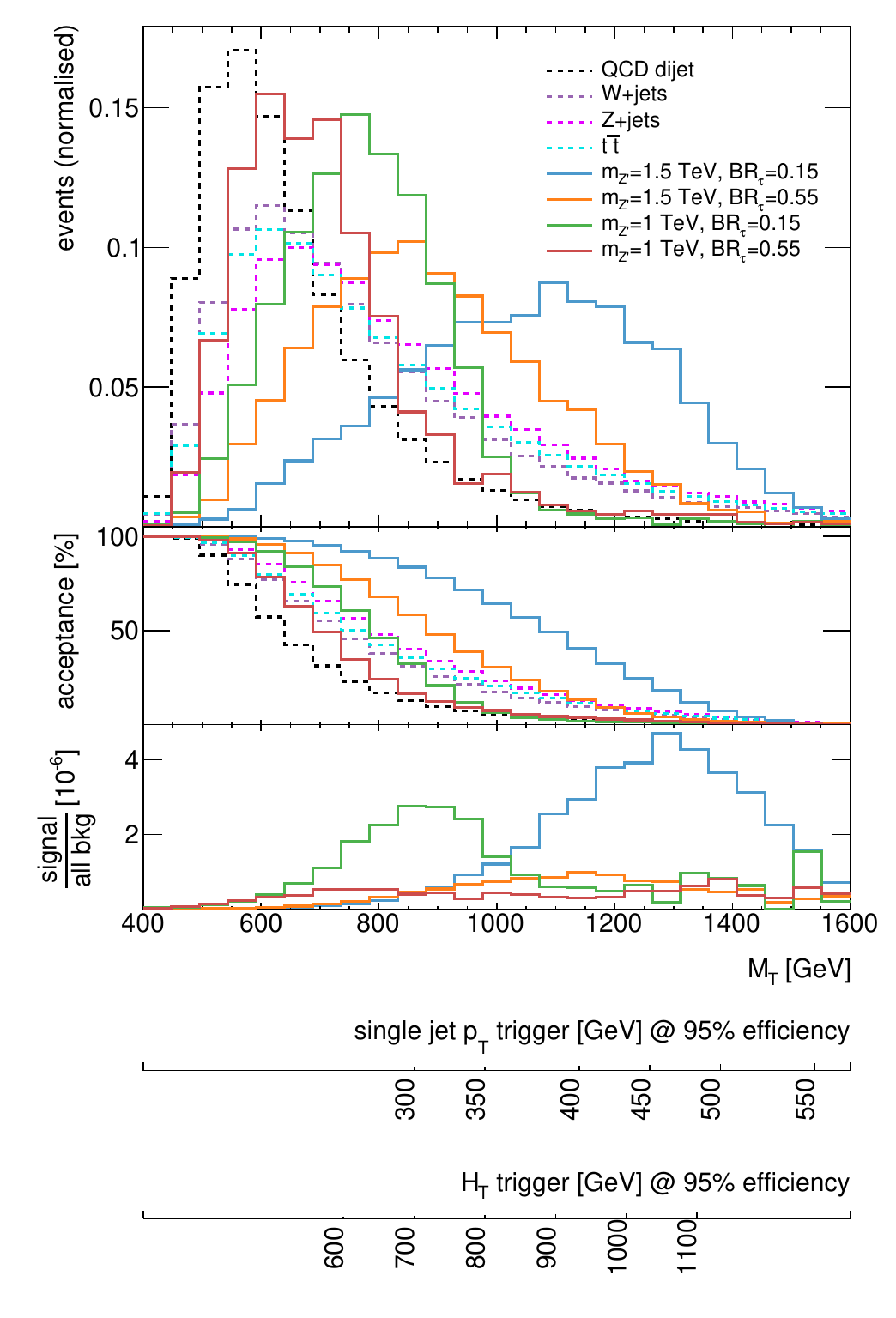}
 \caption{ Signal (solid) and background (dotted) distributions for various $\text{BR}_\tau$ hypotheses, for $M_{Z^\prime}=1$ and $1.5$ TeV.
  Also shown are acceptance, i.e. fraction of events passing a lower threshold on $M_\text{T}$, and the bin-by-bin signal-to-background ratio. For the latter, all backgrounds considered were added together.The distributions in the top panel are normalised to have an integral of one, while for the lower panel the distributions were scaled according to cross-section times branching ratio calculated for $g_u=0.25$.
  All distributions are shown after applying the preselection in the text.
  The axes on the bottom of the plot show the $M_\text{T}$ values at which various trigger selections have been determined to be 95\% efficient in the Delphes simulation with respect to the QCD dijet sample.}
  \label{fig:acceptance}
\end{figure}

A previous CMS search for fully hadronic SVJ signatures \cite{CMS:2021dzg} made it evident that with the HLT $H_\text{T}$ or single jet $p_\text{T}$ trigger thresholds available during Run 2 the possibility to search for low-mass signatures would be very limited.
In the following section several alternative trigger strategies are simulated and evaluated with respect to their effectiveness for the SVJ$\tau$ signal presented in this Letter.
For this purpose, the following event selection based on \cite{CMS:2021dzg} is applied:
\begin{itemize}
    \item Number of good jets: $N(j^\text{AK8})\geq2$, $p_\text{T}(j_{1,2}^\text{AK8})>200$~GeV and $|\eta(j_{1,2}^\text{AK8})|<2.4$
    \item Jet separation: $\Delta\eta(j_1^\text{AK8},j_2^\text{AK8})<1.5$
    \item Jet-$\cancel{E}_{\text{T}}$ separation: $\Delta\phi_{\text{min}}(\cancel{E}_{\text{T}},~j_{1,2}^\text{AK8}) < 0.8$
    \item Veto on mini-isolated leptons: $N_{\mu,e} = 0$
\end{itemize}
\noindent
In \cite{CMS:2021dzg} a criterion on the transverse mass of the two highest $p_\text{T}$ AK8 jets \footnote{It is computed from the 4-vector of the di-jet system $(E_{\text{T},jj},\vec{p}_{\text{T},jj})$ and $(\cancel{E}_{\text{T}},\vec{\cancel{E}_{\text{T}}})$ \cite{Cohen:2015toa}: $M^2_{\text{T}} = (E_{\text{T},jj}+\cancel{E}_{\text{T}})^2 - (\vec{p}_{\text{T},jj}+\vec{\cancel{E}_{\text{T}}})^2$.} $M_{\text{T}}(j_1^\text{AK8},j_2^\text{AK8})>1.5$ TeV is applied in addition to this.
It is driven by the requirement of using a fully efficient trigger.
This selection however limits the sensitivity to semi-visible jets at low mass, especially for SVJ$\tau$ signals, where the presence of neutrinos from $\tau$-leptons decay shift the reconstructed $M_{\text{T}}$ to lower values (see Figure \ref{fig:acceptance})\footnotetext[3]{In order to emulate trigger effects in the Delphes simulation, the selection on $p_\text{T}$ and $H_\text{T}$ was applied to anti-$k_T$ $R=0.4$ jets, while $M_\text{T}$ was calculated using $R=0.8$ jets. As expected, this approach leads to turn-on curves and $M_\text{T}$ cut values that are more optimistic with respect to what is observed in the CMS search. There the full efficiency point for $M_\text{T}= 1.5$ TeV was reached using thresholds of $p_\text{T}=500$ GeV and $H_\text{T}=1.05$ TeV, while for the same thresholds in this study we reach full efficiency for $M_\text{T}=1.2$ TeV.}. 

\noindent 
As a result, a selection $M_\text{T}>1.5$ TeV, following the current trigger strategy for s-channel hadronic semi-visible jets as in \cite{CMS:2021dzg}, discards most of the signal for benchmark points below 2 TeV.\\
\noindent 
The possibility of employing $\tau$ triggers similar to those used by ATLAS and CMS during Run 2 \cite{ATLASTauTrig,CMSTauTrig} was studied as an alternative
(\ref{app:acc_after_preselec}). 
These triggers apply calorimeter and track isolation criteria at different levels of the triggers to select tau candidates, and different options for isolation regions and track thresholds were studied. 
It was found that the isolation requirements impacts the signal acceptance after preselection, especially for small $\text{BR}_\tau$:
The acceptance ranges from $\approx5~\%$ at BR$_\tau=15~\%$ to $\approx35~\%$ at BR$_\tau=70~\%$, with comparable results for all of the studied mass points $0.5\leq M_{Z^\prime}\leq5$ TeV.
The isolation criterion $N_{\text{core: }R_c<0.2}^{\text{tracks: }p_\text{T}>1~\text{GeV}}\leq3$ and $N_{\text{iso: } 0.2<R_i<0.4}^{\text{tracks: } p_\text{T}>1~\text{GeV}}\leq 1$ defined with a core region with radius $R_c=0.2$ was observed to yield generally better acceptances by a factor of $\approx 1.5$ than with a radius of $R_c=0.1$.


\noindent 
The isolation criteria could be loosened using di-$\tau$ triggers, similar to those used in the ATLAS $H\rightarrow\tau\tau$ analysis \cite{ATLAS-CONF-2022-032}, where the isolation is only applied to tau candidates below 25 GeV.
It was observed however 
that such a selection rejects significantly more signal events for low values of $M_{Z^\prime}$, without offering a significant improvement over the single tau trigger.
A future study may investigate whether modifying the trigger isolation criteria for di-tau or dilepton triggers (e.g. introducing variables such as those used in the multivariate analysis presented in section \ref{sec:4}) may reduce trigger rates without dramatically decreasing the signal acceptance.\\
The invisible component of the SVJ$\tau$ signature also suggests the investigation of $\cancel E_\text{T}$ triggers.
However, it was observed
that such a trigger would not reach an acceptable efficiency for the selection of SVJ$\tau$ signal samples with masses below $2$ TeV.\\
\noindent
A topological trigger that makes use of the angle between the two jets was studied and determined to be a more promising approach due to the fact that it takes into consideration the $s-$channel production of the mediator. 
This kind of selection can be applied already at the first trigger level (L1), to reduce hadronic rates prior to the High Level Trigger (HLT). 
The presence of a pair of anti-$k_T$ $R=0.4$ jets with $p_\text{T}>100$ GeV and an angular separation of $\Delta \eta< 2$ was required in combination with the $H_\text{T}$ HLT trigger as specified above.
Comparing the acceptance of QCD background and signal events with and without the additional $\Delta R$ requirement, a larger decrease in the acceptance was observed for the background.
The QCD background acceptance is reduced by about 30\% by imposing the topological requirement on an $H_\text{T}$ trigger threshold of $600$ GeV, and similar reductions are obtained for higher thresholds. 
These studies suggests that topological trigger strategies might aid in lowering the $H_\text{T}$ trigger thresholds.\\
Hadronic trigger thresholds can also be lowered by using alternative data-taking workflows such as Data Parking \cite{CMS-DP-2012-022} (delayed stream in ATLAS \cite{ATLAS:2014ktg}), or Data Scouting \cite{2017520} (Trigger-Level Analysis in ATLAS \cite{PhysRevLett.121.081801}). 
In Run~2, the latter technique only retained global information about the jets reconstructed at the HLT. 
For additional sensitivity to these models using the selections described in the next section, these techniques will need to be supplemented with variables or full event information in the detector region corresponding to the jet.  
\noindent 
In the following analysis, we will use a hypothetical target $M_\text{T}$ cut as low as $800$ GeV for low-mass signals, corresponding to an $H_\text{T}$ trigger threshold of $600$ GeV, according to Figure \ref{fig:acceptance}.  Further studies to demonstrate the feasibility of such a threshold (using e.g. the topological $H_\text{T}$ trigger at L1, in combination with a tau selection at the HLT, or employing alternative data-taking workflows) are required due to the approximations inherent in the simulation setup used for these studies. 


\section{Lepton flavour-based jet tagger}
\label{sec:4}
\begin{figure}[!htbp]
 \centering
 \includegraphics[width=.48\textwidth]{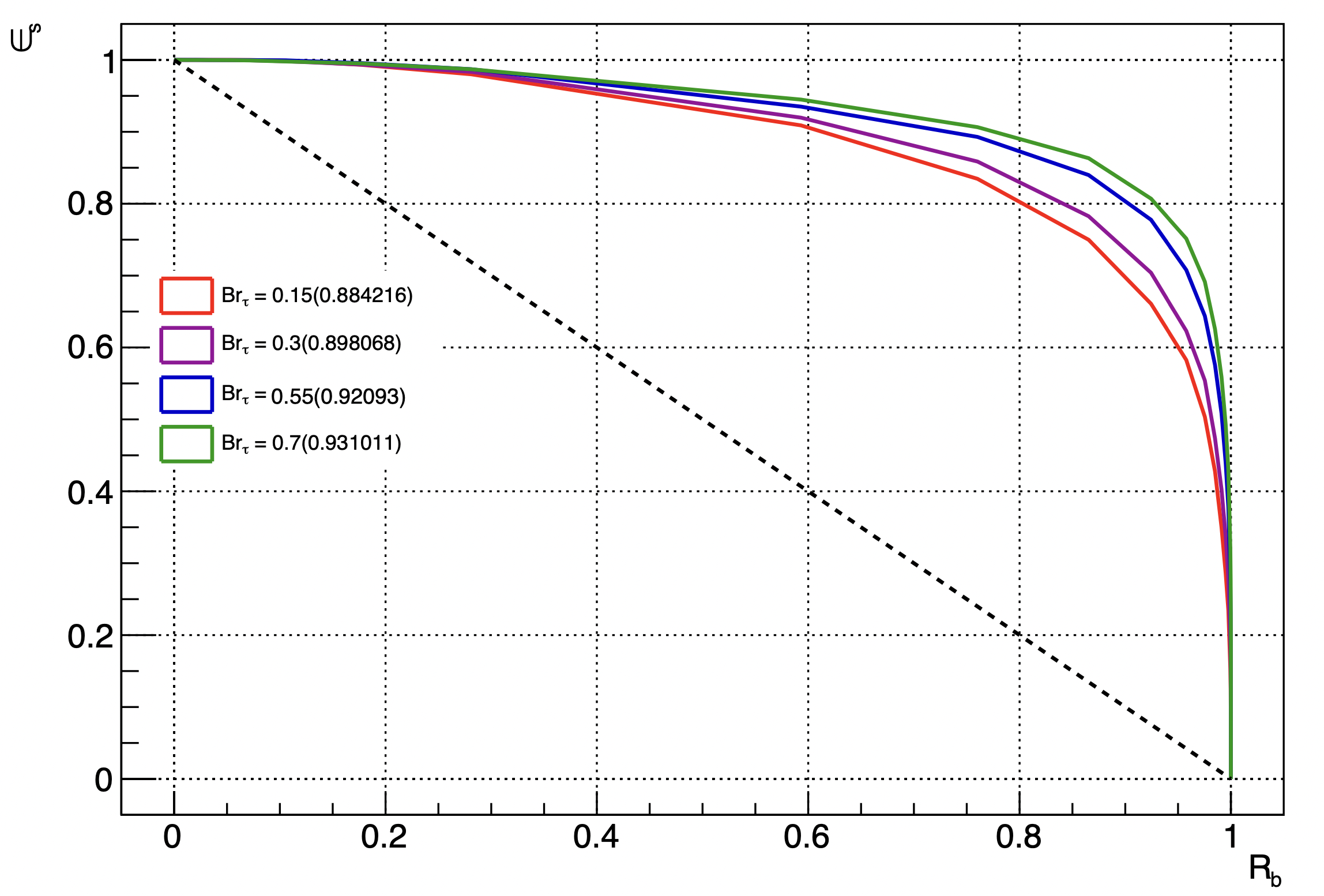}
\caption{Performance of the Boosted Decision Tree jet tagger for $M_{Z'}=3 $ TeV. AUCs are reported in the legend for different values of $\text{BR}_{\tau}$.\label{fig:BDT performance} }
\end{figure}
\begin{figure*}[!tbp]
\centering
\includegraphics[scale=0.42]{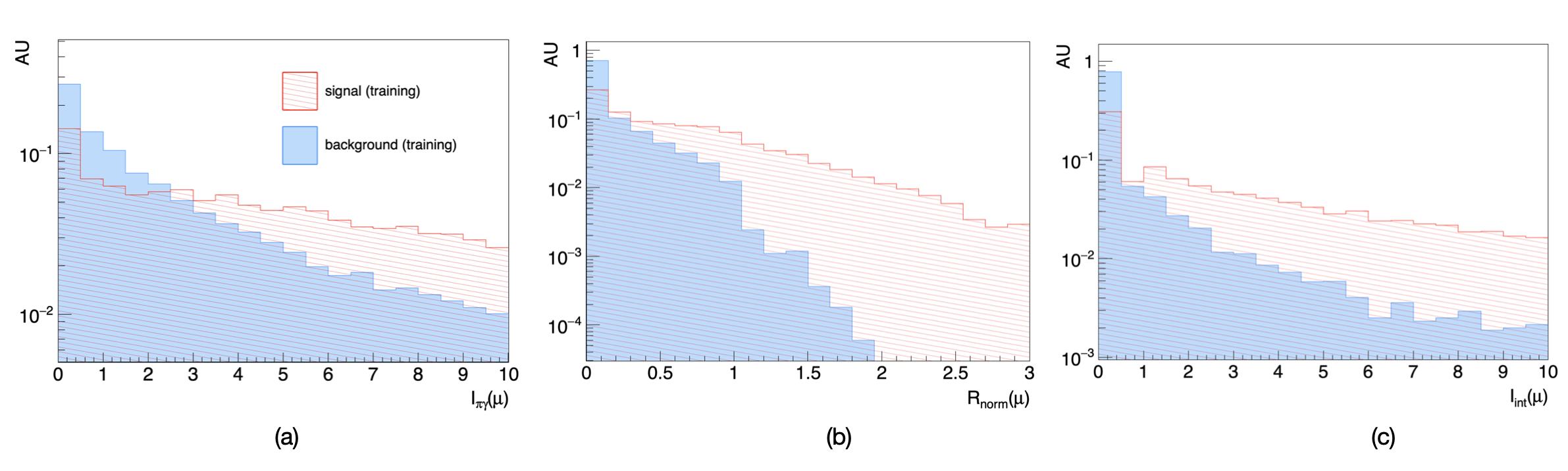}
{\caption{Muon input variables with the highest ranking used for training the jet tagger. \textbf{(a)} $\pi/\gamma$-based isolation, \textbf{(b)} normalised radial distance, \textbf{(c)} inter-isolation. All distributions shown refer to the training sample and are normalised to one. \label{fig:bdt_input_vars}}}
\end{figure*} 
\noindent 
The selection requirements employed by the CMS semi-visible jets search \cite{CMS:2021dzg} rely on event-level variables or basic kinematic properties of the reconstructed jets and missing transverse momentum. Even if this selection rejects the vast majority of the SM background events (Table~\ref{tab:cut_flows_s}, \ref{app:cut_flows}), in the case of relatively large $\text{BR}_{\tau}$, it does not capture the peculiarities of the SVJ$\tau$ signature, with larger lepton multiplicity, additional missing momentum from neutrinos, and larger spread of the jet constituents due to $\tau$ decays. In order to reduce the background further and enhance the sensitivity of the search, we develop here a novel jet tagging algorithm leveraging the specific features of signal jets. In this work, we propose a set of new variables able to capture the atypical lepton content in the signal jets due to $\tau$ lepton decays.  
Indeed, since the signal is enriched in $\tau$ leptons, and $\sim 35\%$ of them is expected to decay to electrons and muons, we expect a relevant contribution of light leptons  enriching the jet signature. Compared to the SVJ-$\ell$ signature presented in~\cite{Cazzaniga:2022hxl}, the leptonic fraction in the jets is lower, and the $p_\text{T}$ distribution for electrons and muons is expected to be much softer due to the presence of neutrinos in $\tau$ decays. The signature addressed here is therefore more challenging. 
An adequate background reduction requires a good exploitation of multiple properties of the signal leptons.
In our lepton flavour-based jet tagger we exploit 3 main features of electrons and muons (labelled $\ell = e, \mu$) in signal jets: energy and momentum flow, spatial distribution and isolation. All the variables proposed have to be intended to be computed using  candidates from the Particle-Flow algorithm \cite{CMS:2017yfk,ATLAS:2017ghe}, which allows to combine the information from all the sub-detectors to identify particles \footnote{In Delphes only a simplified version of the Particle-Flow algorithm is implemented \cite{deFavereau:2013fsa}.}. In particular, the QED radiation from electrons bending in the detector magnetic field is resummed in the definition of the object itself via a Super-Clustering algorithm \cite{CMS:2020uim,ATLAS:2019qmc}. The same algorithm also  allows to properly reconstruct the original photons which converted into $e^{+}e^{-}$ pairs due to interactions with the detector material \cite{CMS:2015myp}. For the first category of variables energy fractions ($f(\ell)$) are used together with the average lepton  $p_{\text{T}}$ normalised by the jet transverse momentum ($p_{\text{T,norm}}(\ell)$). The second category of variables includes the radial distribution of electrons and muons with respect to the jet axis normalised by the jet radius ($\text{R}_{norm}(\ell)$). Finally, the third class of variables includes the inter-isolation ($\text{I}_{\text{int}}(\ell)$) proposed in~\cite{Cazzaniga:2022hxl}, as well as a new isolation variable: $\text{I}_{\pi \gamma}(\ell)$. The former variable, namely the inter-isolation, quantifies the leptonic activity around a given lepton in a jet  summing in the isolation cone only electrons and muons in the whole event. On the other hand, the $\text{I}_{\pi \gamma}(\ell)$ isolation is built summing the momenta of all charged pions and photons reconstructed in the event within a cone of given radius centred around the candidate lepton itself associated to a given jet, and then normalising to the lepton $p_{\text{T}}$. This variable aims to capture $\pi_v \to \tau(h)\tau(\ell)$ when the dark hadron $\pi_v$ is boosted enough and the $\tau(h)$ products overlap with the lepton from $\tau(\ell)$. Both isolation variables are calculated introducing an infrared regulator via a  minimum transverse momentum requirement $p_{\text{T}} = 0.1$ GeV for the particles summed in the isolation cone. It has been checked that the variable $\text{I}_{\text{int}}(\ell)$, despite being collinear unsafe with respect to QED radiation, it is not significantly altered by including photons in its definition. Both the isolation variables proposed have been found to be uncorrelated, and both powerful, thus it is beneficial to exploit both to obtain the best performance for the jet-tagger.\\ The input variables sensitive to the multiplicity of jet constituents are deliberately excluded, as this property is strongly correlated with the details of the dark sector. \\ \\
These variables are used as input to a Boosted Decision Tree (BDT). The list of input variables used with their definition is in Table \ref{tab:bdt_inputs} (\ref{app:jet_tagger}). The BDT is trained using the TMVA machine learning package \cite{Hocker:2007ht} with the adaptive boosting technique. The inputs are the two highest $p_{\text{T}}$ jets from simulated signal and background samples, with the variables described above computed for each jet. To improve the performance of the classifier, its hyperparameters have been optimised for a benchmark signal point at $M_{Z'} = 3$ TeV mixing different $\text{BR}_{\tau}$ hypotheses. Each signal sample is weighted equally. Moreover, in order to lower correlation to leptons multiplicity, which is highly dependent on the details of the model, only signal and background jets with non-zero leptonic content have been considered for the training.\\
The BDT exhibits strong rejection of jets from the SM background processes, including heavy-flavour jets,
while preserving a significant amount of signal jets. The receiver operator characteristic curves (ROC)  for a benchmark mass point of $M_{Z'} = 3$ TeV and different values of $\text{BR}_{\tau}$ considering all background jets are shown in Figure \ref{fig:BDT performance}. It is expected that the performance of the tagger decreases with smaller values of $\text{BR}_{\tau}$, since the chosen input variables aim to exploit the enhanced fraction of electrons and muons due to decay processes $\tau \to \ell \bar{\nu}_{\ell} \nu_{\tau}$ occurring inside the jet. In Figure \ref{fig:AUC-table} (\ref{app:jet_tagger}), it is reported the evolution of the Area Under the Curve (AUC), representing the performance of the jet tagger, for a grid of signal hypothesis parameters.  The three most important variables in the BDT, thus those with the highest signal-versus-background overlap integral separation as defined in \cite{Hocker:2007ht}, are found to be $\text{I}_{\text{int}}(\ell)$, $R_{norm}(\ell)$ and $I_{\pi \gamma}(\ell)$. The distributions for such high ranking variables  for muons are shown in Figure \ref{fig:bdt_input_vars}. We report in Table \ref{tab:bdt_inputs} (\ref{app:jet_tagger}) the ranking of input variables used in the BDT. \\
We choose a working point (WP) corresponding to a threshold of 0.05 on the discriminator output. Jets with a discriminator value higher than 0.05 are labelled as semi-visible.  This threshold has been chosen to maximise the significance $s/\sqrt{s + b}$, where $s$ and $b$ are the expected number of signal and background events passing the BDT cut, respectively. The distributions of the BDT response for signal and background with the relative optimised selection are reported in Figure \ref{fig:BDT-performance} (\ref{app:jet_tagger}). At this WP, the BDT rejects $\sim 97\%$ of simulated background jets, while correctly classifying $\sim 80\%, 84\%, 90\%, 93\%$ of jets from the benchmark signal models at $M_{Z'}=3$ TeV for $\text{BR}_{\tau}=0.15, 0.3, 0.55, 0.7$, respectively. \\
It has been verified that the set of variables presented here present higher discrimination power with respect to  the jet-substructure variables used in~\cite{CMS:2021dzg}.
\begin{figure*}[!htbp]
  \centering
  \includegraphics[width=1.0\textwidth]{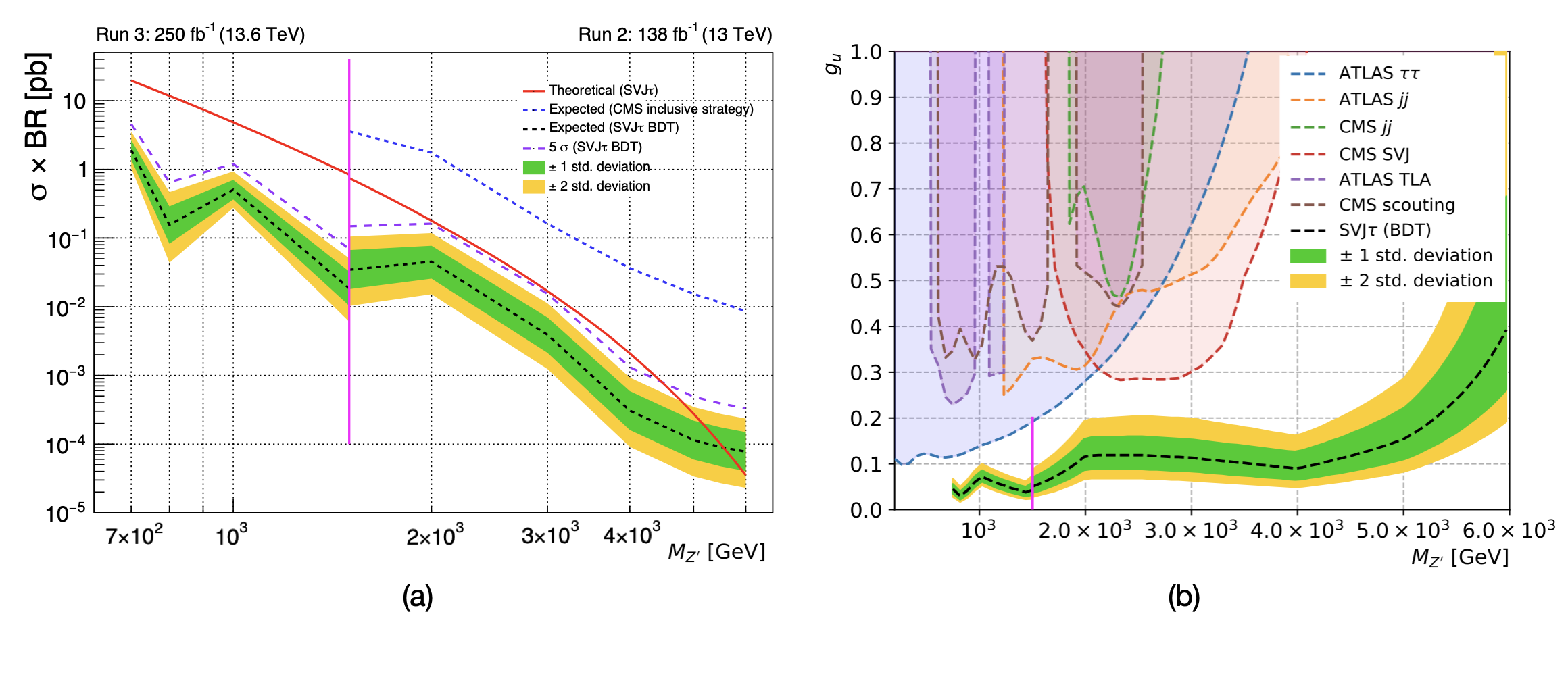}
\caption{\textbf{(a)} Expected limits on $\sigma \times BR(Z' \to q_v \bar{q}_v)$ for the signal benchmark with $r_{\text{inv}} = 0.3$ and $\text{BR}_{\tau} = 0.55$. The low-mass region ($700 \leq M_{Z'} \leq 1500$ GeV) accessible in Run~3 when requiring $M_{\text{T}} > 800$ GeV is separated by a vertical magenta line from the high mass search region ($M_{Z'} \geq 1500$ GeV) accessible using the same triggers as in \cite{CMS:2021dzg} for Run~2. The expected exclusion reach of the fully hadronic SVJ inclusive analysis for the SVJ$\tau$ for the SVJ$\tau$ signature (dashed blue line) is compared with the proposed SVJ$\tau$ BDT-based analysis (black dashed line), exploiting the BDT jet tagger introduced in section \ref{sec:4}. The theoretical cross-section (red line) has been computed for a benchmark $g_{u} = 0.25$ and $g_{q_v} = 0.6$. The purple line corresponds to the expected $5\sigma$ discovery line. \textbf{(b)} Current constraints and expected limits of the BDT-based analysis as a function of the mass of the mediator $M_{Z'}$ for $\text{BR}_\tau = 0.55$, $r_{\text{inv}}=0.3$ and $g_{q_v} = 0.6$. The black line corresponds to the SVJ$\tau$ BDT-based analysis.}
\label{fig:limit_result}
\end{figure*}
\section{Results}
\label{sec:5}
We report in this section the sensitivity and exclusion reach for the SVJ$\tau$ signal, both in the case of a high mass search, like the one reported by the CMS collaboration for Run~2~\cite{CMS:2021dzg}, thus keeping the same requirement $M_\text{T} > 1500$ GeV, and with a new trigger strategy for Run~3 requiring a reference selection $M_{\text{T}} > 800$ GeV to probe lower $M_{Z'}$ masses. We have estimated the expected exclusion limit at 95\% confidence level (CL) for $\sigma \times \text{BR}$ for different $M_{Z'}$ and $\text{BR}_{\tau}$ hypotheses performing a binned likelihood template fit of the $M_{\text{T}}$ spectrum and using the modified frequentist approach $\text{CL}_{\text{s}}$ in the asymptotic approximation~\cite{Junk:1999kv,Read:2002hq,Cowan:2010js}. 
The sensitivity reach has been estimated for events selected according to the strategy defined in section~\ref{sec:4}. We have combined two categories of events according to the jet-level decision of the tagger at the optimal WP (as introduced in section~\ref{sec:5}): $0$-tag category, where no jets are tagged as semi-visible, and $1/2$-tag category where at least one jet is tagged as semi-visible. \\
We consider a benchmark point with $r_{\text{inv}}= 0.3$ and $\text{BR}_{\tau} = 0.55$. For the high mass search, the LHC with full Run~2 collision data ($\sqrt{s}=13$ TeV, $\mathcal{L}_{int} = $ 138 $\text{fb}^{-1}$) and following the BDT-based approach proposed here is expected to claim the discovery (exclusion) of an hypothetical $Z'$ boson with SM couplings as described in section \ref{sec:Model} up to masses of \mbox{$\sim 4.5$ TeV ($5.5$ TeV)} for a benchmark $g_u = 0.25$. We show that, the search could be able to probe mediator masses down  to 700~GeV with full Run~3 LHC collision data ($\sqrt{s}=13.6$ TeV, $\mathcal{L}_{int} = $ 250 $\text{fb}^{-1}$ ).  More details are shown in Figure \ref{fig:limit_result}\textcolor{blue}{.a}. The sensitivity to the signal diminishes with higher values of $\text{BR}_{\tau}$ due to the decreasing $M_\text{T}$ resolution. Furthermore,  Figure \ref{fig:limit_result}\textcolor{blue}{.a}, shows how the inclusive CMS SVJ search~\cite{CMS:2021dzg} has no sensitivity to SVJ$\tau$ signals, while the jet tagger proposed in this Letter allows probing values of the coupling $g_{u}$ even smaller than the benchmark 0.25, for the entire $M_{Z'}$ mass range tested.\\
 Figure \ref{fig:limit_result}\textcolor{blue}{.b} shows the expected limits of the BDT-based analysis on the coupling $g_u$ for the same benchmark, together 
 with previous experimental constraints. Six searches are included: the dijet search by ATLAS~\cite{ATLAS:2019fgd}, the dijet search by CMS~\cite{CMS:2019gwf}, the ditau search by ATLAS~\cite{ATLAS:2017eiz}, the SVJ search by CMS~\cite{CMS:2021dzg}, the trigger-object-level analysis (TLA) search by ATLAS~\cite{ATLAS:2018qto} and the dijet data-scouting search by CMS~\cite{CMS:2018mgb}. Dijet searches only include jets originating from SM quarks, and their constraints are conservative as some SVJ would also have passed the selection cuts (a detailed study is beyond the scope of this work). In general, if the $Z'$ also decays to dark quarks, the branching ratios to SM jets and taus will be reduced, weakening the constraints from dijet and ditau searches. As can be seen, the proposed strategy vastly outperforms previous searches.\footnote{The expected limits are so strong that dark bound states could be forced to decay displaced. Prompt decays could still be obtained by the inclusion of other mediators, see e.g. \cite{Cazzaniga:2022hxl}.}
\section{Discussion and conclusions}
\label{sec:6}
In our study, we have investigated the possibility to access lower $Z'$ boson masses by lowering the $M_\text{T}$ requirement. We have found that topological triggers based on $H_\text{T}$, $\tau$ triggers for high tau branching ratios, and alternative data-taking techniques such as data scouting or parking can be promising to achieve an increased sensitivity with respect to the current trigger strategy. \\
Finally, we provided the expected sensitivity reach and exclusion limits performing a bump hunt in the dijet transverse mass both in the case of a high mass search fulfilling the  CMS trigger requirements for the Run 2 SVJ search~\cite{CMS:2021dzg}, and also for a lower mass search exploiting a different trigger strategy for Run 3 LHC data taking. For the high mass search, the BDT-based analysis can claim the discovery (exclusion) of a $Z'$ mediator as defined in section \ref{sec:Model} up to masses of 4.5 TeV (5.5 TeV) with full Run2 collision data when the fraction of unstable dark hadrons decaying to $\tau$ lepton pairs is around $50\%$ and with a benchmark coupling $g_u = 0.25$. For the same benchmark parameter space, we estimate that our BDT-based strategy is also capable to probe $Z'$ boson masses down to 700 GeV when lowering the $M_\text{T}$ requirement down to 800 GeV. 

\begin{acknowledgements}
 We are grateful to M. Strassler for his useful suggestions and comments. We thank F. Eble and R. Seidita for providing substantial developments to the code used to perform the study presented here, and revising the manuscript providing useful comments. We also thank the participants of the first SVJ Workshop held in Zurich in July 2022, for useful discussions. C. Cazzaniga and A. de Cosa  are supported by the Swiss National Science Fundation (SNFS) under the SNSF Eccellenza program. H. Beauchesne is supported by the National Science and Technology Council, the Ministry of Education (Higher Education Sprout Project NTU-111L104022), and the National Center for Theoretical Sciences of Taiwan. G.G.d.C. is supported by the INFN Iniziativa Specifica Theoretical Astroparticle Physics (TAsP) and by the Frascati National Laboratories (LNF) through a Cabibbo Fellowship, call 2019.
C. Doglioni and T. Fitschen's research is part of a project that has received funding from the European Research Council under the European Union’s Horizon 2020 research and innovation program (grant agreement 101002463).
\end{acknowledgements}

\appendix

\section{Event selection}
\label{app:cut_flows}
The event selection  used in this work is shown in the first column of Table \ref{tab:cut_flows_s}. This is based on a previous CMS study~\cite{CMS:2021dzg}. The last three columns shows the signal efficiencies for different values of BR$_\tau$ and a benchmark mass point $M_{Z'}=3$ TeV.

\begin{table}[htb]
	\centering
    \renewcommand{\arraystretch}{1}
	\begin{tabular}{c c c c }\hline 
	\multirow{2}{2em}{Selection} &
	    \multicolumn{3}{c}{Signal efficiency ($\%$)} \\
		$\quad \quad \quad \quad \quad \quad \quad \quad \quad \quad \quad \quad\text{BR}_{\tau}:$& 0.7 & 0.55 & 0.15 \\\hline \hline
		$N(AK8 jets) \geq 2$           &71.8    &   76.5  &85.7   \\\hline
		$\Delta\eta (j_1,j_2)  < 1.5$  &51.5    &   53.5  &56.6   \\\hline
		$M_{\text{T}} > 1500$ GeV                &16.2    &   25.5   &42.7     \\\hline
		$R_{\text{T}}>$  0.15             & 10     &   14.3     &20.2     \\\hline
		$\Delta\phi_{\text{min}}(\cancel{E}_{\text{T}},j) <$  0.8  &9.3     &   13.6    &19.5  \\\hline
		$N_{\mu,e}$ = 0 &5.7     &   9.2      &17.3 \\\hline
		\end{tabular}
		\caption{CMS SVJ search \cite{CMS:2021dzg} selections applied to SVJ$\tau$ signal for a benchmark mass point $M_{Z'}=3 $ TeV.}
		\label{tab:cut_flows_s}
\end{table}
\renewcommand{\arraystretch}{0.2}

\section{Acceptance for $\tau$ Triggers}\label{app:acc_after_preselec}

\begin{figure}[!htbp]
 \centering
 \includegraphics[width=.5\textwidth]{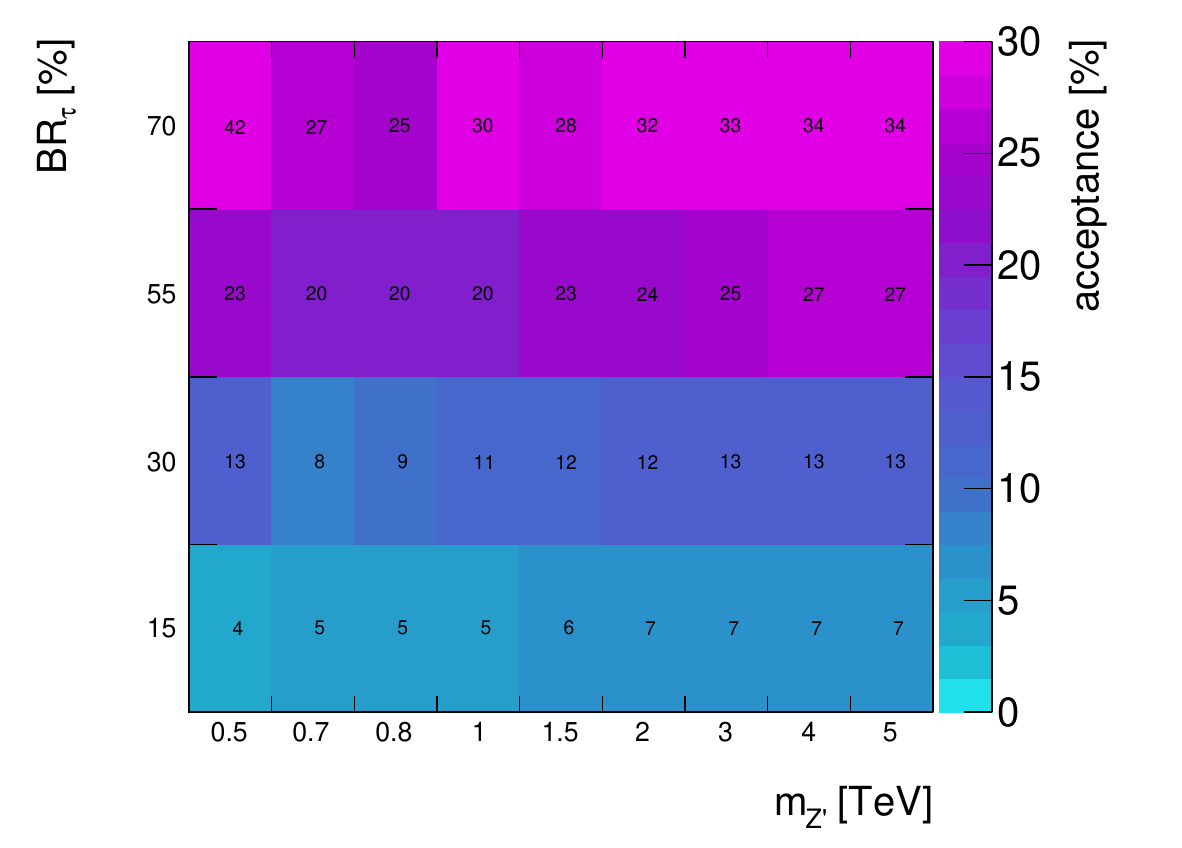}\\
 \includegraphics[width=.5\textwidth]{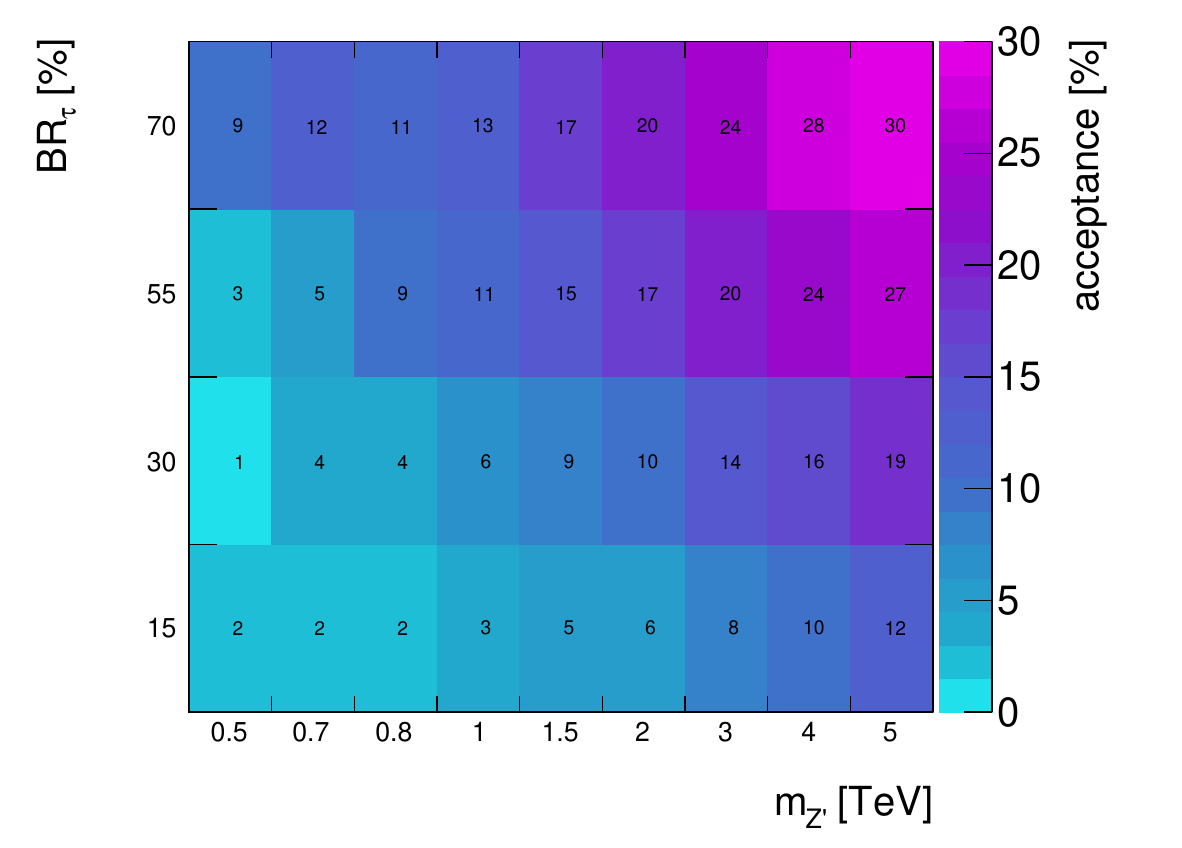}
 \caption{
  \label{fig:tauTrig}
  [top]: Signal acceptance for events fulfilling the criteron of at least one particle level $\tau$ fulfilling the single $\tau$ trigger isolation criterion of $1\leq N_\text{core}^\text{tracks}\leq3$ and $1\leq N_\text{iso}^\text{tracks}$ with $R_\text{core}=0.1$ [bottom]: Signal acceptance for events fulfilling the di-$\tau$ trigger requirement of $p_\text{T}(\tau^\text{lead})>35$ GeV and  $p_\text{T}(\tau^\text{sublead})>25$ GeV at particle level for various mass points $M_{Z^\prime}$ as well as branching ratios $\text{BR}_\tau$.
  The preselection as defined in \ref{sec:3} has been applied.
 }
\end{figure}

Figure~\ref{fig:tauTrig} shows the signal acceptance after the isolation criterion and the preselection as specified in Sec. \ref{sec:3} for different BR$_\tau$ and $M_{Z'}$ with respect to the single- and di-$\tau$ triggers specified in the same section.

\section{Jet tagger input variables and performance}\label{app:jet_tagger}
\begin{table}[htb]
	\centering
    \renewcommand{\arraystretch}{1}
	\begin{tabular}{c c c c }\hline 
        $\text{Rank}$ & $\text{Variable}$ & $\text{Definition}$   \\\hline \hline \\[-1em]
		1  &  $\text{I}_{\text{int}}(\mu)$   & $\sum^{\Delta \text{R} < \text{R}_{\text{iso}}}_{\ell \neq \mu}p_{\text{T}}(\ell)/p_{\text{T}}(\mu)$        \\\hline \\[-1em]
		2  &  $\text{R}_{\text{norm}}(\mu)$   &  $\sum_{\ell \in \text{jet}} \Delta \text{R}({\mu,\ell}) / \text{R}$  \\\hline \\[-1em]
        3  &  $\text{I}_{\pi \gamma}(\mu)$  & $\sum^{\Delta \text{R} < \text{R}_{\text{iso}}}_{i \in \{\pi^{\pm},\gamma \}\in \text{jet}}p_\text{T}(i)/p_{\text{T}}(\mu)$     \\\hline \\[-1em]
        4  &   $\text{R}_{\text{norm}}(e)$  & $\sum_{\ell \in \text{jet}} \Delta \text{R}(e,\ell) / \text{R}$     \\\hline \\[-1em]
        5  &  $\text{I}_{\text{int}}(e)$  &  $\sum^{\Delta \text{R} < \text{R}_{\text{iso}}}_{\ell \neq e}p_\text{T}(\ell)/p_{\text{T}}(e)$   \\\hline \\[-1em]
        6  &  $f(e)$   &  $\sum_{e \in \text{jet}} p_{\text{T}}(e)/\sum_{i \in \text{jet} } p_{\text{T}}(i)$    \\\hline \\[-1em]
        7  & $\text{I}_{\pi \gamma}(e)$   & $\sum^{\Delta \text{R} < \text{R}_{\text{iso}}}_{i \in \{\pi^{\pm},\gamma \}\in \text{jet}}p_\text{T}(i)/p_{\text{T}}(e)$     \\\hline \\[-1em]    
        8  &  $p_{\text{T},\text{norm}}(\mu)$   & $p_{\text{T}}(\mu)/p_{\text{T}}(\text{jet})$     \\\hline \\[-1em]
        9  & $f(\mu)$    &  $\sum_{\mu \in \text{jet}} p_{\text{T}}(\mu)/\sum_{i \in \text{jet} } p_{\text{T}}(\text{i})$   \\\hline \\[-1em]
		\end{tabular}
		\caption{Input variables to the jet tagger ordered by decreasing separation (as defined in \cite{Hocker:2007ht}). In the third column are reported the definitions of the variables used with the following conventions: $\text{R}$ is the jet radius, $\ell$ is a generic lepton (electron or muon), $\text{R}_{\text{iso}}$ is the isolation cone radius (for $\text{I}_{\text{int}}(\ell)$ is fixed to $\text{R}_{\text{iso}} = 0.5$, while for $\text{I}_{\pi \gamma}(\ell)$ is fixed to $\text{R}_{\text{iso}} = 0.3$. Both $\text{I}_{\text{int}}(\ell)$ and $\text{I}_{\pi \gamma}(\ell)$ are calculated with a  minimum transverse momentum requirement $p_{\text{T}} = 0.1$ GeV for the particles summed in the isolation cone.}
		\label{tab:bdt_inputs}
\end{table}
\renewcommand{\arraystretch}{1.0}

\begin{figure}[!htbp]
 \centering
 \includegraphics[width=.4\textwidth]{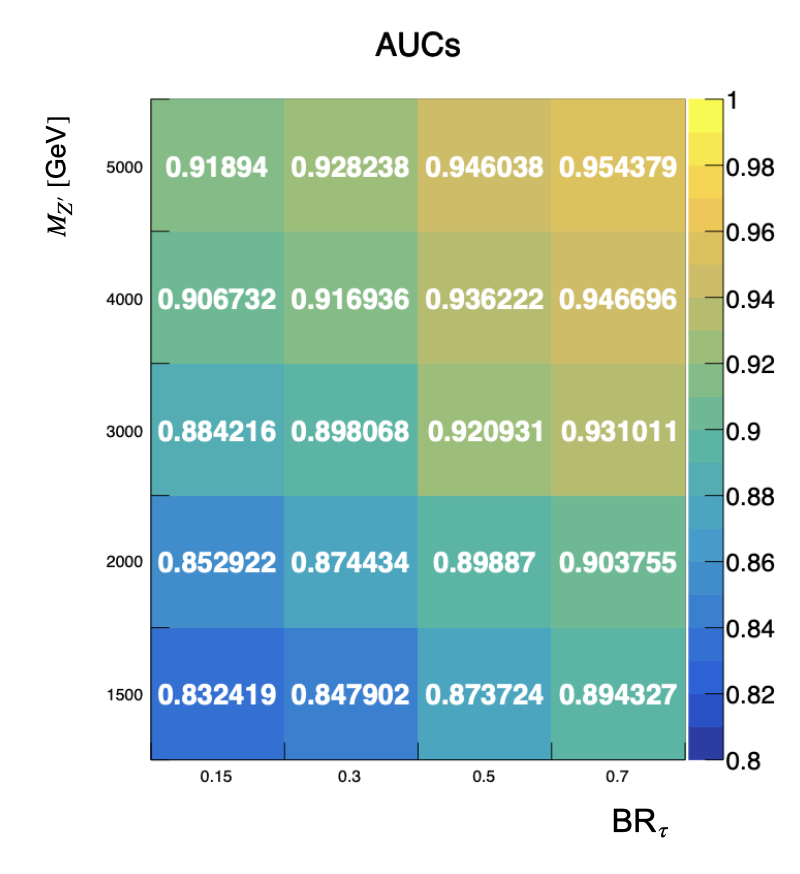}
 \caption{
  \label{fig:AUC-table}   
 Performance of the jet tagger for different signal hypothesis changing the parameters $M_{Z'}$ and $\text{BR}_\tau$. The colour gradient and numbers represents the AUC of the ROC curve for the given signal hypothesis.  
 }
\end{figure}

\begin{figure}[!htbp]
 \centering
 \includegraphics[width=.5\textwidth]{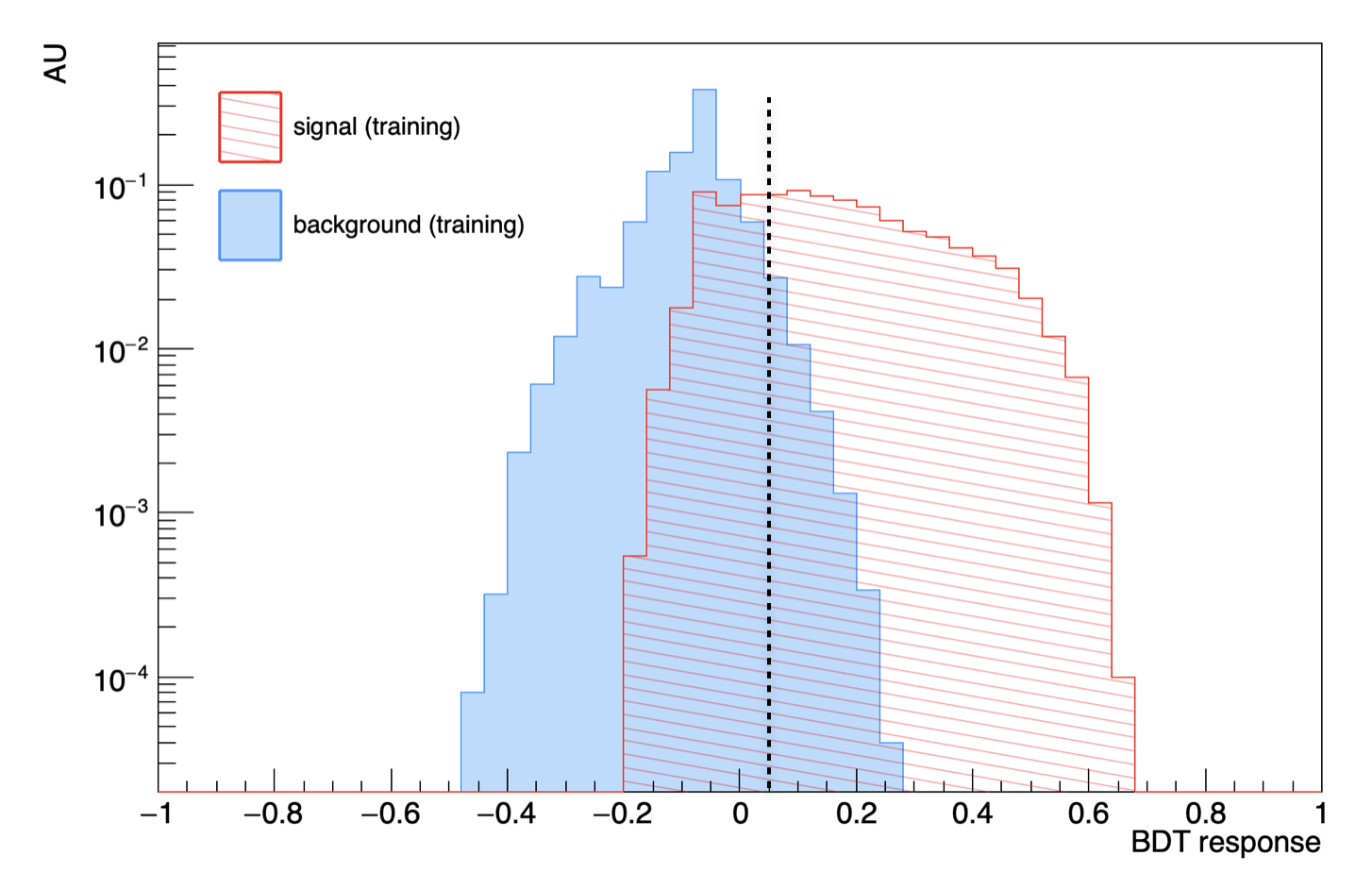}
 \caption{
  \label{fig:BDT-performance}
 BDT response for signal (red) and background (blue) training datasets. The vertical black dashed line represents the best cut value chosen maximising $s/\sqrt{s+b}$.    
 }
\end{figure}

\bibliographystyle{spphys}       
\bibliography{bibl}   

\end{document}